\documentclass[aps,prb,superscriptaddress,amsmath,amssymb,onecolumn]{revtex4-1}

\usepackage{graphicx}
\usepackage{hyperref}
\usepackage{epstopdf}
\usepackage{bm,bbm}

\begin{document}

\title{Inter-adsorbate forces and coherent scattering in helium spin-echo experiments.}

\author{David J. Ward}
\email{djw77@cam.ac.uk}
\affiliation{Cavendish Laboratory, J.J. Thomson Avenue, Cambridge CB3 0HE, United Kingdom}
\author{Arjun Raghavan}
\affiliation{Cavendish Laboratory, J.J. Thomson Avenue, Cambridge CB3 0HE, United Kingdom}
\affiliation{Department of Physics, University of Illinois at Urbana-Champaign - 1110 W. Green St., Urbana, IL 61801}
\author{Anton Tamt\"{o}gl}
\affiliation{Cavendish Laboratory, J.J. Thomson Avenue, Cambridge CB3 0HE, United Kingdom}
\affiliation{Institute of Experimental Physics, Graz University of Technology, Petersgasse 16, 8010 Graz, Austria}
\author{Andrew P. Jardine}
\author{Emanuel Bahn}
\affiliation{Cavendish Laboratory, J.J. Thomson Avenue, Cambridge CB3 0HE, United Kingdom}
\author{John Ellis}
\affiliation{Cavendish Laboratory, J.J. Thomson Avenue, Cambridge CB3 0HE, United Kingdom}
\author{Salvador Miret-Art\'es}
\affiliation{CSIC Institute of Fundamental Physics, Serrano 123, Madrid 28006, Spain}
\author{William Allison}
\affiliation{Cavendish Laboratory, J.J. Thomson Avenue, Cambridge CB3 0HE, United Kingdom}

\begin{abstract}
In studies of dynamical systems, helium atoms scatter coherently from an ensemble of adsorbates as they diffuse on the surface. The results give information on the co-operative behaviour of interacting adsorbates and thus include the effects of both adsorbate-substrate and adsorbate-adsorbate interactions. Here, we discuss a method to disentangle the effects of interactions between adsorbates from those with the substrate. The result gives an approximation to observations that would be obtained if the scattering was incoherent. Information from the experiment can therefore be used to distinguish more clearly between long-range inter-adsorbate forces and the short range effects arising from the local lattice potential and associated thermal excitations. The method is discussed in the context of a system with strong inter-adsorbate interactions, sodium atoms diffusing on a copper (111) surface.
\end{abstract}

\maketitle

\section{Introduction}

Helium atom scattering is well established as a sensitive probe of surface processes and particularly for studies of the dynamics of pristine and adsorbate covered surfaces\cite{Benedek2018}. Atoms are known to scatter strongly and coherently from the electron density at a surface. Thus, the technique is particularly powerful in analysing vibrations that have weight at, or near, the surface\cite{Benedek2020,Benedek2020b,Tamtogl2018d}. The resulting inelastic scattering gives both the energy and wavelength dependence of the excitations. Helium atoms are also sensitive to changes in the electron density as adsorbed atoms and molecules move on an otherwise flat surface\cite{Alexandrowicz2007,Jardine2009a,Fratesi2008}.  Scattering under these circumstances generates quasi-elastic intensity, where time correlations in the intensity reflect the time-dependence of motion in the target.  The present work is concerned with the diffusion of adsorbates and, in particular, the analysis of quasi-elastic scattering in helium spin-echo measurements when strong inter-adsorbate forces are present.

The scattering of any wave from a dynamical assembly of particles encodes information on the factors that underlie motion within the assembly. The difficulties of extracting that information from fluctuations in the scattered intensity have been recognised since the earliest experiments of dynamical light-scattering\cite{YAN1989277}. In the case of a system in equilibrium, it was argued that the time-scales of thermal excitation, inter-particle forces and the time-resolution in the experiment all have an effect on the observations\cite{Pusey1975}. The degree of coherence in the scattering process also has an important contribution to the outcome of an experiment\cite{Leitner2011}.  

In a classical, kinematic approximation, the amplitude scattered from a moving particle, $j$, with position, $\mathbf{r}_j(t)$, will depend on the momentum transfer, $\Delta \mathbf{K}$, and time, $t$, as follows
\begin{equation}
 A_j \left( \Delta \mathbf{K} , t \right) = f(\Delta\mathbf{K}) \exp \left[ {-\mathbbm{i} \Delta \mathbf{K}\cdot{\mathbf{r}_j(t)}} \right] \; ,
 \label{eq:AjKt}
\end{equation} 
where the amplitude form-factor, $f(\Delta \mathbf{K})$, depends on the spatial distribution of the scattering centre.  In neutron scattering the moving particles are point-like and the form-factor is independent of $\Delta \mathbf{K}$ to a good approximation\cite{Fouquet2010,Calvo2016,Jones2016}. We are concerned with the motion of adsorbates on a surface and thus the charge distribution from which the helium atoms scatter has a form-factor that must be retained if, as here, the intensity distribution is important to the analysis.

When the scattering is coherent then the amplitudes from the individual adsorbates sum to give a total amplitude
\begin{equation}
 A(\Delta \mathbf{K} , t)=\frac{1}{N}\sum_j A_j=\frac{f(\Delta\mathbf{K})}{N}\sum_j  \exp \left[ {-\mathbbm{i} \Delta \mathbf{K}\cdot{\mathbf{r}_j(t)}} \right] \; ,
 \label{eq:AcohKt}
\end{equation} 
where we have chosen to normalise the amplitude by dividing by the total number, $N$, of adsorbates.  The intensity, $\langle A \cdot A^* \rangle $, for coherent scattering is
\begin{equation}
\begin{split}
 I_{coh}(\Delta \mathbf{K},t) & = \left\langle A(\Delta \mathbf{K},\tau) \cdot A^*(\Delta \mathbf{K},t+\tau) \right\rangle \\
& =\frac{\mathopen |f(\Delta\mathbf{K}) \mathclose|^2}{N^2} \sum_{jj^{\prime}} \left\langle \exp{ \left[-\mathbbm{i} \Delta \mathbf{K} \cdot \mathbf{r}_j(\tau) \right] } \exp{ \left[ \mathbbm{i} \Delta \mathbf{K}\cdot{\mathbf{r}_{j^{\prime}}(t+\tau)} \right] }  \right\rangle\; .
 \end{split}
 \label{eq:cohintensity}
\end{equation}
The angle brackets indicate an ensemble average, which is equivalent to an average over the time variable, $\tau$, when the dynamics are ergodic.

When the scattering is incoherent the intensity from each adsorbate,
\begin{equation}
\begin{split}
 I_j(\Delta \mathbf{K},t)&  = \langle A_j(\Delta \mathbf{K},\tau) \cdot A_j^*(\Delta \mathbf{K},t+\tau) \rangle \\
 & =\mathopen |f(\Delta\mathbf{K}) \mathclose|^2 \left\langle \exp{ \left[ -\mathbbm{i} \Delta \mathbf{K} \cdot{\mathbf{r}_j} (\tau) \right] } \exp{ \left[ \mathbbm{i} \Delta \mathbf{K}\cdot{\mathbf{r}_j(t+\tau)} \right] } \right\rangle
 \end{split}
 \label{eq:IjKt}
\end{equation} 
is summed to give the total intensity for incoherent scattering
\begin{equation}
\begin{split}
 I_{incoh}(\Delta \mathbf{K},t)  & =\frac{1}{N} \sum_{j} I_j \\
 & =\frac{\mathopen |f(\Delta\mathbf{K}) \mathclose|^2}{N} \sum_{j} \left\langle \exp{ \left[-\mathbbm{i} \Delta \mathbf{K} \cdot \mathbf{r}_{j}(\tau) \right] } \exp{ \left[ \mathbbm{i} \Delta \mathbf{K} \cdot \mathbf{r}_j(t+\tau) \right] } \right\rangle\; .
 \end{split}
 \label{eq:incohintensity}
\end{equation}
Here the normalisation ensures that, when all adsorbates scatter in phase, the coherent and incoherent intensities are equal.

The incoherent intensity (\autoref{eq:incohintensity}) is determined entirely by the self-correlation of individual scattering centres, $j$, whereas the coherent intensity ( \autoref{eq:cohintensity}) includes correlations between all pairs of particles, $j$ and $j^{\prime}$. Cases where the motion is co-operative and where correlations between particles are important would clearly generate differences in the two measures of intensity fluctuation. Since co-operative motion requires some degree of interaction between the adsorbates,  differences between the coherent and incoherent intensities reflect the nature of inter-particle forces. Incoherent scattering is regarded as having a more intuitive interpretation since it is indicative of the local adsorbate-substrate potential and the thermal excitations that control the dynamics\cite{Leitner2011}. In contrast,  coherent scattering will show, in addition, the effects of interactions.

Helium scattering is inherently a coherent scattering technique. However, there is a direct advantage in having access to both coherent and incoherent scattering intensities when analysing data from surface systems. In a typical diffusion study the aim is first to establish the energy landscape on which the particles move. The landscape is defined by the principal adsorption sites and the transition states for diffusion. Incoherent scattering provides a simple method that helps to distinguish the effects of the local energy landscape from long-range interactions that are generated by forces acting between the adsorbates. In that way, simple dynamical models, such as idealised hopping \cite{Chudley1961, Tuddenham2010} can be used to generate a first-cut model of the landscape before inter-adsorbate forces are considered. A more complete analysis can then proceed by analysing the strength and range of the forces that contribute to the coherent scattering. In this way a better, self-consistent description of the experimental data can be obtained.

The relationship between coherent and incoherent lineshapes is known qualitatively as de Gennes narrowing\cite{Degennes1959}, though the interpretation in terms of cooperative behaviour remains a topic for debate\cite{Wu2018}. A quantitative relationship between the incoherent and coherent correlation functions has only been established for approximate systems such as site-to-site hopping motion of weakly interacting particles in three-dimensional space (3-D)\cite{SINHA198851,Leitner2009,Leitner2011}. Here, we explore the validity of the quantitative approach for the study of strongly interacting adsorbates, in a 2-D system, where the effects of correlated motion dominate the scattering.

The approach we take in the present work is first to establish a `typical' system of strongly-correlated adsorbates.  Here we consider sodium atoms moving on a copper (111) surface, for which both a model landscape and an interaction model are available\cite{Rittmeyer2016}. We use simulations based on the Langevin molecular-dynamics framework to deduce the coherent and incoherent scattering intensities, upon the assumption of point scatterers. The analysis indicates that 3-D models\cite{SINHA198851,Leitner2011} have validity when applied to strongly correlated motion in this 2-D system.  Experimental results from the Na/Cu(111) system are then analysed and we demonstrate that a suitable form-factor for scattering can be constructed. The form-factor then allows us to obtain an incoherent scattering intensity from the measurements of coherent scattering.

\section{Analysis of coherent and incoherent scattering}

Intensity correlation functions, as given in \autoref{eq:cohintensity} and \ref{eq:incohintensity}, are known as Intermediate Scattering Functions (ISF) in the neutron scattering literature. The interpretation of these correlation functions forms the basis of the quasi-elastic scattering technique. 
For unconfined lateral diffusion of the adsorbates, the long-time limit of the correlation function is assumed to decay exponentially, $I (\Delta \mathbf{K},t) \sim \exp [-\alpha(\Delta \mathbf{K}, t)]$\cite{Townsend2018a}. In general, therefore, the intensity correlation function has a characteristic `lineshape' that may be written as
\begin{equation}
\begin{split}
 I(\Delta \mathbf{K},t) &= \mathopen |f(\Delta\mathbf{K})\mathclose|^2 \, B(\Delta \mathbf{K},t) \, \exp  \left[ -\alpha(\Delta \mathbf{K}, t) \right] \\
 & = b(\Delta \mathbf{K},t) \, \exp \left[ -\alpha(\Delta \mathbf{K}, t) \right] \; ,
 \end{split}
 \label{eq:Lineshape}
\end{equation}
where $\alpha$ is the dephasing rate. The prefactor, $b(\Delta \mathbf{K},t)$, decays to a constant value in the limit $t \rightarrow \infty $\cite{Townsend2018a}. At shorter times the prefactor $b(\Delta \mathbf{K},t)$ may have a complicated structure. For example, it may contain multiple exponential decays\cite{Tuddenham2010} and, at very small times it has a Gaussian time dependence that describes ballistic motion\cite{Vega2004,Guantes2004}.

The quantitative relationship between coherent and incoherent lineshapes has been studied theoretically, in the context of neutron scattering\cite{SINHA198851,Leitner2011} and dynamical light scattering\cite{Pusey1975}. These studies use approximations to make the algebra tractable. Typically they involve the interpretation of weakly interacting systems in 3-D. Examples include the diffusion of dilute interstitial particles, or diffusion in alloys. Usually, the motion is assumed to occur in the absence of an external potential\cite{Pusey1975}, or on a well-defined lattice in the regime where hops are the dominant dynamical process, giving a quasi-elastic scattering lineshape that is a simple exponential\cite{Clapp1966,SINHA198851,Leitner2009,Leitner2011}. 
In the case of hopping, the incoherent dephasing rate, $\alpha_{incoh}$, is then assumed to have the form derived by Chudley and Elliott \cite{Chudley1961}
\begin{equation}
\alpha_{incoh}(\Delta\textbf{K})=\sum_n \tfrac{1}{T} \big[ 1 - \exp ( \mathbbm{i}\Delta\textbf{K}\cdot\textbf{R}_n ) \big]
\label{eq:ChudleyElliott}
\end{equation}
for a residence time, $T$, with $n$ jump-sites having jump vectors, $\mathbf{R}_n$. 

The dephasing rates for coherent and incoherent scattering encode the time dependence of the motion and, when interactions are present, they  will have a different dependence on $\Delta \mathbf{K}$.  In the limit of weak interactions between scattering centres, the relationship between the two dephasing rates is known\cite{SINHA198851,Leitner2011}. Derivations use a self-consistent field calculation within linear-response theory\cite{SINHA198851}, or obtain a similar result using transition-state theory\cite{Leitner2011}. In both derivations, the dephasing rates for coherent and incoherent scattering are related by the prefactor for coherent scattering, $B_{coh}(\Delta\textbf{K},t=0)$, so that
\begin{equation}
\begin{split}
 \alpha_{incoh}(\Delta\textbf{K}) & =\alpha_{coh}(\Delta\textbf{K}) \, B_{coh}(\Delta\textbf{K},t=0) \\
 & = \alpha_{coh}( \Delta\textbf{K} ) \, \frac{ b_{coh}(\Delta\textbf{K},t=0) }{ |f(\Delta\textbf{K})|^2 } \; .
 \end{split}
 \label{eq:incohalpha}
\end{equation}
Sinha and Ross\cite{SINHA198851} were the first to include the motion of spatially extended objects having a defined form-factor, $|f(\Delta\textbf{K})|^2$, and hence derive equation \eqref{eq:incohalpha} in the form given. In their case the extended object was a lattice distortion surrounding a moving interstitial atom; however, their argument is equally applicable to scattering from the distribution of electronic charge surrounding a moving adsorbate, as in the present work. The prefactor for coherent scattering, $b_{coh}(\Delta\textbf{K},t=0)$, is known as either the quasi-elastic contribution to the structure factor\cite{SINHA198851} or the intensity due to short range order\cite{Leitner2011}. 

Equation \eqref{eq:incohalpha} is significant as it shows that measurements of coherent scattering can be used, in principle, to deduce the dephasing rates for incoherent scattering.  
The method outlined above is widely used in weakly-interacting 3-D systems but the approximations are untested in the context systems with strong spatial correlation, such as diffusion in 2-D, and for a strongly scattering probe such as helium atoms. In the present paper we explore the value of \autoref{eq:incohalpha} in the context of surface systems with strong correlations in the motion. We show that incoherently scattered intensity can be deduced from measurements of coherent helium scattering, at a good level of approximation.

\section{Numerical simulations of coherent and incoherent scattering}

The motion of adsorbates on a periodic surface may be reproduced with various forms of molecular dynamics simulations. In the present work we are concerned with the calculation of scattering from the adsorbates and, for that purpose, the Langevin, or Generalised-Langevin framework provides a convenient and well established method to understand the motion\cite{Ala-Nissila2002,Miret-Artes2005,Jardine2009b,Townsend2018,Avidor2019}. Here, the dynamical coordinates of the adsorbates are treated explicitly while the substrate interactions are represented by an adiabatic potential-energy surface. Thermal excitation is represented by a combination of random forces and an appropriate frictional force and it is possible to include an explicit description of inter-adsorbate interactions. 

Differences in the calculated intensity correlations  using the Langevin or Generalised-Langevin equations depend on the frequency spectrum of the thermal noise; however, those differences disappear as the correlation-time extends beyond any correlations in the noise spectrum\cite{Townsend2018}. For these reasons we adopt the computationally more efficient Langevin approach. The equation of motion for the dynamical coordinates, $\mathbf{r}_j$, is
\begin{equation}
 m\ddot{\mathbf{r}}_j=-\nabla{V}(\mathbf{r}_j)-\eta{m\dot{\mathbf{r}}_j}+\epsilon(t)+\sum_{j\neq{j'}}F(\mathbf{r}_{j'}-\mathbf{r}_j) \; ,
 \label{eq:langevineqn}
\end{equation}
where the adsorbates with mass $m$, interact with the substrate through an adiabatic potential-energy surface, $V(\mathbf{r}_j)$, and are subject to a stochastic force, $\epsilon(t)$, with a white-noise spectrum.  The stochastic force is balanced, on average, by a velocity dependant retarding force $-\eta{m\dot{\mathbf{r}_i}}$. Pairwise interactions between adsorbates, $j$ and $j'$, are introduced by the force $F(\mathbf{r}_j'-\mathbf{r}_j)$.  By including the pairwise adsorbate forces explicitly, we obtain an accurate description of correlated motion that goes beyond stochastic models of interacting adsorbates\cite{MarinezCasadoJPCM2007,Martinez-CasadoPRL2007}.

\begin{figure}[htbp]
\centering
\includegraphics[width=0.42\textwidth]{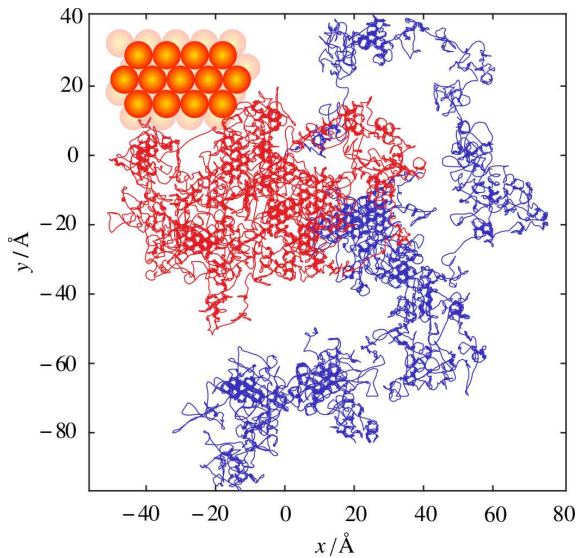}
\caption{Sample trajectories for two representative adsorbates (red and blue) out of a simulation of 500 interacting species. The simulation data was collected during a total run time of 6500\,ps. Parameters, such as the adiabatic potential, coverage ($\Theta=0.025$ monolayer) and friction ($\eta=0.43\,$ps$^{-1}$) are the same as those reported in Ref\cite{Rittmeyer2016} and correspond to values that describe the experimental data discussed below. Both fcc and hcp sites have a local minimum in the potential and the system temperature is such that single hops between adjacent sites dominate the motion. Occasionally, the trajectories exhibit long hops due to the low value of the friction $\eta$. Inset shows the real space atomic structure of the substrate, measurements and simulations reported are performed on the $]11\bar{2}]$ direction which is aligned with the $y$ axis.}
\label{fig:fig1}
\end{figure}

The equations of motion may be integrated for discrete time steps, $\delta{t}$, building up a `trajectory', $\mathbf{r}_i(t)$, for each adsorbate. Sample trajectories for two atoms taken from a simulation are shown in \autoref{fig:fig1}. The simulation includes 500 interacting sodium atoms on a copper[111] surface at a temperature of $155\,\mbox{K}$. The adiabatic potential, coverage of $\Theta=0.025$\,monolayer defined with respect to the saturation coverage and friction $\eta=0.43\,$ps$^{-1}$ are taken from\cite{Rittmeyer2016} and correspond to values that describe the experimental data discussed below. Kohn-Lau \cite{Kohn1976} dipole-dipole inter-adsorbate forces are included, parameterised according to \cite{Rittmeyer2016}. The trajectories in \autoref{fig:fig1} map out a honeycomb structure on which hopping between sites takes place. A honeycomb trajectory arises when both hcp and fcc sites of the (111) surface act as adsorption sites\cite{Lechner2013a}. Adsorbates spend a significant time at one adsorption site until gaining sufficient energy to overcome the energy barrier between sites. Occasional long hops are evident in the trajectories. In these instances the adsorbate traverses two or more barriers, before becoming trapped again at a particular adsorption site. The fraction of the long hops, relative to single jumps, depends on the friction\cite{Ala-Nissila2002} and those that are evident in \autoref{fig:fig1} are consistent with the low friction used in the simulation.  Co-operative motion due to the effects of interactions between adsorbates cannot easily be seen from a superficial inspection of trajectories such as in \autoref{fig:fig1}, but the effects are clearly apparent in the correlation functions we discuss below.

We calculate the coherent and the incoherent scattering intensities defined in \autoref{eq:cohintensity} and \ref{eq:incohintensity} using the relevant amplitudes, $A(\Delta \mathbf{K},t)$ according to \eqref{eq:AcohKt} and $A_j(\Delta \mathbf{K},t)$ according to \eqref{eq:AjKt} respectively, where the phase factor, $a_j(\Delta \mathbf{K},t)= \exp [-\mathbbm{i} \Delta \mathbf{K} \cdot \mathbf{r}_j(t) ]$, is constructed from the trajectory, $r(t)$.  Its temporal Fourier transform
\begin{equation}
 a_j(\Delta \mathbf{K},\omega)=\mathcal{F} \left[ a_j(\Delta \mathbf{K} , t) \right] \; ,
 \label{eq:AjKw}
\end{equation}
is a useful and efficient tool in calculating the intensity correlation\cite{Ellis1997}. The convolution theorem gives
\begin{equation}
\begin{split}
 I_j(\Delta \mathbf{K},t) &= \left\langle A_j(\Delta \mathbf{K},\tau) \cdot A_j^*(\Delta \mathbf{K},t+\tau) \right\rangle \\
 &= \mathopen |f(\Delta\mathbf{K})\mathclose|^2 \, \mathcal{F}^{-1} \left[ a_j(\Delta \mathbf{K},\omega) \cdot a_j^*(\Delta \mathbf{K},\omega) \right]  \; ,
 \label{eq:Ij}
  \end{split}
\end{equation}
for the correlations in the intensity scattered from a single adsorbate, $j$. The incoherent intensity follows from an average over trajectories, as shown in \autoref{eq:incohintensity}. For coherent scattering, the scattering amplitudes for all trajectories are summed (\autoref{eq:AcohKt}) before the Fourier transform,
\begin{equation}
 a(\mathbf{K},\omega)= \mathcal{F} \bigg[ \sum_{j} a_j (\mathbf{K},t) \bigg] \; ,
 \label{eq:AKw}
\end{equation} 
which leads to the intensity for coherent scattering
\begin{equation}
\begin{split}
 I_{coh}(\Delta \mathbf{K},t) &= \frac{1}{N^2} \big\langle A(\Delta \mathbf{K},\tau) \cdot A^*(\Delta \mathbf{K},t+\tau) \big\rangle  \\
 &= \frac{ \mathopen |f(\Delta\mathbf{K})\mathclose|^2 } {N^2} \, \mathcal{F}^{-1} \bigg[ a(\Delta \mathbf{K},\omega) \cdot  a^*(\Delta \mathbf{K},\omega) \bigg] \; .
 \label{eq:Icoh}
 \end{split}
\end{equation}

In general the intensity correlation functions are more complex than the single exponential decay assumed in the derivation of  \autoref{eq:incohalpha} and complete analytic forms are only known for a limited number of simple systems\cite{Vega2004,Guantes2004,Townsend2018}. Fortunately, it is not necessary to know the analytic form of the amplitude in order to explore the validity of \autoref{eq:incohalpha}.
We do, however, need to determine the two key quantities, $\alpha_{coh}(\Delta \mathbf{K})$, and, $b_{coh}(\Delta \mathbf{K},t=0)$.

In the work below we use a similar procedure to analyse the intensity correlation functions from the numerical simulations and the experimental data, taking the form-factor, $\mathopen |f(\Delta\mathbf{K})\mathclose|^2 = 1$, in the simulations. First, the dephasing rates, $\alpha(\Delta\mathbf{K})$, for coherent and incoherent scattering are extracted by fitting the simulated intensities, $I_{incoh}$ and $I_{coh}$, at long times to a decaying exponential $b_o \exp \left [ -\alpha(\Delta\mathbf{K},t) \right]$ \cite{WardThesis} (see also Supplementary Information of Rittmeyer et.al.\cite{Rittmeyer2016}).
Second, we extract the `short time scale' contribution to the structure factor using a Gaussian, $b_1 \exp \left[ -(t/\sigma)^2 \right]$, where the width parameter, $\sigma$, approximates the ballistic motion and any remaining terms, at short times. The quasi-elastic contribution to the structure factor from  \autoref{eq:Lineshape} is then given by, 
\begin{equation}
 b_{coh}(\Delta \mathbf{K},t=0)=b_o + b_1 \; .
 \label{eq:QEamplitude}
\end{equation}

\begin{figure}[htbp]
\centering
\includegraphics[width=0.42\textwidth]{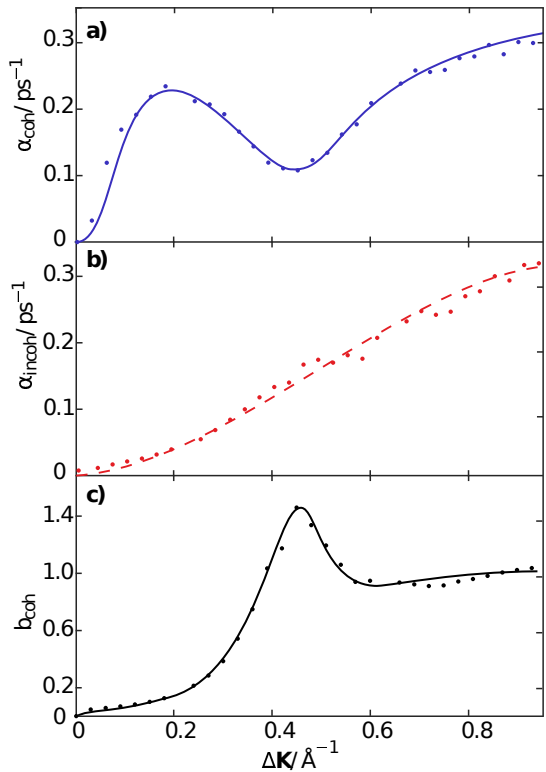}
\caption{Analysis of trajectory data for scattering along the $[11\bar{2}]$ direction ($y$-axis in inset to figure \ref{fig:fig1}). Panel (a) shows the coherent dephasing rates, $\alpha_{coh}(\Delta \mathbf{K})$ calculated using parameters according to \cite{Rittmeyer2016}. The dephasing rates for coherent scattering, shown as blue data points with a line to guide the eye, are extracted from the long-time limit of the intensity correlation function (see text). Panel (b) shows incoherent dephasing rates, $\alpha_{incoh}(\Delta \mathbf{K})$, as a dashed red line.  The rates are calculated in the same way as in (a) but using \autoref{eq:incohintensity}. The red data points are deduced from results in panels (a) and (c), with $f(\Delta \mathbf{K})=1$. The black points in panel (c) show the quasi-elastic contribution to the structure factor, $b_{coh}(\Delta \mathbf{K},t=0)$ (see text) with the solid line to guide the eye. Values that deviate from unity indicate the effects of inter-adsorbate interactions on the resulting dephasing rates.
\label{fig:fig2}}
\end{figure}

\autoref{fig:fig2} shows results from simulations of Na/Cu(111) using trajectories such as those in \autoref{fig:fig1}.
Blue data points (\autoref{fig:fig2}(a)) give the dephasing rate for coherent scattering as a function of momentum transfer in the $[11\bar{2}]$ direction. The results are characteristic of strong repulsive forces between the adsorbates\cite{AlexandrowiczNaCu2006} that are  evident in the pronounced maximum and subsequent dip observed for $\Delta \mathbf{K} < 0.5\,\mbox{\AA}^{-1}$. Scatter in the data points arises from the statistical uncertainty of the simulations.

\autoref{fig:fig2}(b) shows the corresponding dephasing rates for incoherent scattering. In this case the results, deduced from  \autoref{eq:incohintensity}, are shown by the red line. Note that the statistical variation is much less than in panel (a) as the incoherent intensities are averaged over all trajectories in addition to the averaging over repeated simulation runs which also takes place in the calculation of the coherent intensities. The data points in the middle panel are calculated from  \autoref{eq:incohalpha} using values of $\alpha_{coh}$ in panel (a) and values of the quasi-elastic contribution to the structure factor, $b_{coh}(\Delta \mathbf{K},t=0)$ shown in \autoref{fig:fig2}(c). The main feature in panel (c) is the dramatic decrease in $b_{coh}(\Delta \mathbf{K},t=0)$ as $\Delta \mathbf{K}$ approaches zero and a similar effect is evident in the experiment (see later). The strong inter-adsorbate interactions maintain the separation of the adsorbates and stabilise structures where interference in the scattered amplitudes from neighbouring adsorbates interfere destructively, giving rise to the low scattered intensity at small $\Delta \mathbf{K}$. The converse is evident in the peak between $0.4-0.5\,\mbox{\AA}^{-1}$, where the same structures tend to scatter constructively and the corresponding scattered intensity is higher.

The excellent agreement between the two estimates of the incoherent dephasing rates (red points and solid line in In \autoref{fig:fig2}(b)) supports both the analytic model\cite{SINHA198851}, \autoref{eq:incohalpha} and our method of analysis.

For the analysis leading to \autoref{fig:fig2} we have taken the adsorbates to be point scatterers and the corresponding form-factor for scattering $f(\Delta \mathbf{K})=1$ in \autoref{eq:incohalpha}.  We now turn to the analysis of experiment and, in particular, the determination of the experimental form-factor.

\begin{figure}[htbp]
\centering
\includegraphics[width=0.42\textwidth]{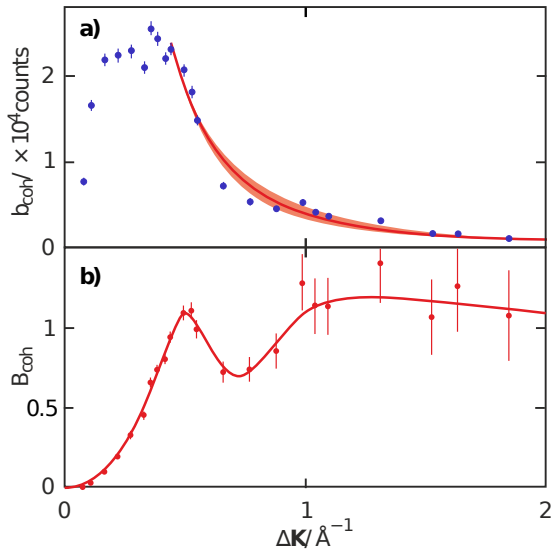}
\caption{Panel (a) shows the measured quasi-elastic intensity, $b_{coh}(\Delta \mathbf{K},t=0)$ as blue data points with estimated statistical uncertainty. The red curve shows a fit to the intensity for $\Delta \mathbf{K} > 0.44\,\mbox{\AA}^{-1}$, where the uncertainty in the exponent, $n$, is indicated by the shaded area. (b) Amplitude of the quasielastic lineshape $B_{coh}(\Delta \mathbf{K},t=0) = b_{coh}(\Delta \mathbf{K},t=0)  / \mathopen |f(\Delta\mathbf{K})\mathclose|^2$ deduced from the data in panel (a) as red data points. The estimates of uncertainty are constructed from the respective uncertainties in the measurements and in $n$, treated as statistical variables. The red line in (b) is provided as a guide to the eye.}
\label{fig:fig3}
\end{figure}

\section{Experimental measurements and the form-factor}
There are significant differences between intensity correlation functions from experimental and from numerical simulation. In the experiment there are contributions from inelastic scattering, due to substrate phonons\cite{Benedek2018}, and purely elastic scattering from static features on the surface contributes to the total intensity.  In addition, the form-factor for scattering from the mobile species, $|f(\Delta\textbf{K})|^2$, must be extracted from the $\Delta \mathbf{K}$ dependence of the scattered intensity.

Contributions from inelastic scattering are removed by a Fourier filter, applied in the frequency domain\cite{WardThesis,Rittmeyer2016}, while scattering from static structures is accounted for by subtracting a constant term so that the intensity has the same form as equation \ref{eq:Lineshape} and $ I(\Delta \mathbf{K},t) \rightarrow 0$ as $ t \rightarrow \infty$.

The coherent intensity can then be analysed in the same way as the simulations (see above). In order to extract an approximate dephasing rate for incoherent scattering according to \autoref{eq:incohalpha}, we need a corresponding approximation to the $\Delta \mathbf{K}$ dependence of the form-factor. Earlier works, such as measurements using time-of-flight methods\cite{Ellis1999}, have been analysed on the basis of a power law to approximate the form-factor
\begin{equation}
 \mathopen |f(\Delta\mathbf{K})\mathclose|^2= C \,  \Delta\textbf{K} ^{-n} \; .
 \label{eq:Formfactor}
\end{equation}
Ellis \textit{et al.}\cite{Ellis1999} found the exponent to be, $n=3$ in the case of Xe atoms on a Pt(111) surface. However, the value of $n$ will depend on the nature of the scattering object.  For example, a value $n=2$ at small values of $\Delta \mathbf{K}$, is predicted for scattering from a 1-D object such as a step\cite{Lahee1987}.

In the present analysis, we treat $n$ as a free parameter and determine the value that best describes the $\Delta \mathbf{K}$ dependence of the measured intensity at large momentum transfers, where $B_{coh}(\Delta \mathbf{K},t=0)$ is approximately constant. \autoref{fig:fig3}(a) shows the measured quasi-elastic intensity, $b_{coh}(\Delta \mathbf{K},t=0)$, as blue data points together with the best-fit to the form-factor, $\mathopen |f(\Delta \mathbf{K})\mathclose|^2$ according to \eqref{eq:Formfactor} (red curve). A value of $n = 2.2 \pm 0.2$ describes the trend in the data for $0.44 <  \Delta \mathbf{K} <2\,\mbox{\AA}^{-1}$. At smaller values of $\Delta \mathbf{K}$ the intensity deviates markedly from $\Delta \mathbf{K}^{-n}$ and we attribute the effect to the expected decrease in quasi-elastic intensity noted in \autoref{fig:fig2}(c), above.

\autoref{fig:fig3}(b) shows the amplitude of the quasi-elastic lineshape, $B_{coh}(\Delta \mathbf{K},t=0) = b_{coh}(\Delta \mathbf{K},t=0) / \mathopen |f(\Delta\mathbf{K})\mathclose|^2$, derived from  the data in panel (a).
 The result shows the same features as observed in the trajectory simulations (\autoref{fig:fig2}(c), above) namely, a reduction in quasi-elastic intensity as $\Delta \mathbf{K} \rightarrow 0$, followed by a maximum corresponding to diffraction from quasi-static structures. The fact that these features emerge strongly from such a simple model for the form-factor ( \autoref{eq:Formfactor}) suggests that the procedure is robust.

\begin{figure}[htbp]
\centering
\includegraphics[width=0.42\textwidth]{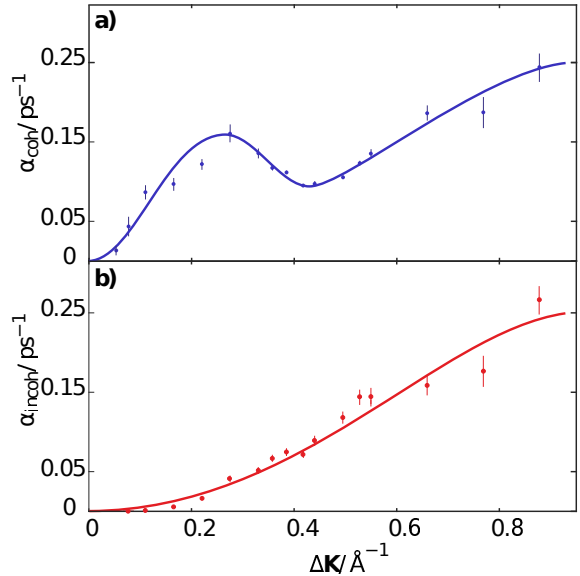}
\caption{Top panel (a) shows the dephasing rate $\alpha (\Delta \mathbf{K})$ for coherent scattering in blue. In the lower panel (b) red points show the dephasing rate for incoherent scattering obtained by scaling the blue data points by $B_{coh}(\Delta \mathbf{K},t=0) = b_{coh}(\Delta \mathbf{K},t=0)  / \mathopen |f(\Delta\mathbf{K})\mathclose|^2$ according to \autoref{eq:incohalpha}. The lines are drawn as a guide to the eye.}
\label{fig:fig8}
\end{figure}

Figure \ref{fig:fig8} shows the coherent and incoherent dephasing rates obtained from coherent scattering data. Blue points in panel (a) are coherent dephasing rates derived from the time-dependence of the experimental data while the red points in lower panel (b) show incoherent rates deduced using  \autoref{eq:incohalpha} and the amplitude of the quasi-elastic lineshape, $B_{coh}(\Delta \mathbf{K},t=0)$, shown in \autoref{fig:fig3}(b). The differences between the red and blue data are as expected. In particular, the shape of the incoherent data (red points) is clearly quadratic at small values of $\Delta \mathbf{K}$ indicating diffusive behaviour in the absence of correlated motion, as would be expected for hopping according to \autoref{eq:ChudleyElliott}.

\section{Conclusions}
The results presented above show that it is possible to remove the influence of inter-adsorbate correlations leading to data equivalent to an incoherent scattering experiment. The data can then be analysed using simple models to give an approximation to the energy landscape before including inter-adsorbate interactions in a more complete analysis.

Our simulations use a molecular-dynamics approach, which includes both inter-cell and intra-cell motion.  However, the analysis would be identical for a simulation using Monte-Carlo methods where the trajectories for each adsorbate are generated by random hops on a specified lattice\cite{Tamtogl2020}.  The resulting lineshape (equation \ref{eq:Lineshape}) is simpler to analyse, since intra-cell motion is absent, but the results should be the same, as long as the Monte-Carlo algorithm generates the correct statistical occupancy of sites\cite{Leitner2011}.

The form-factor in helium scattering is relatively little studied and it is a significant challenge, if equation \ref{eq:incohalpha} is to be applied more widely in quasi-elastic scattering experiments with helium atoms. In the present case, variations in the scattered intensity can be attributed to the form-factor at large $\Delta\mathbf{K}$ and to a variation in the quasi-elastic contribution to the structure factor at low $\Delta\mathbf{K}$. Furthermore the form-factor is well described by a simple power law in dependence of $\Delta\mathbf{K}$. For systems having greater complexity in the dynamics, such as the motion of molecules, the analysis may be more difficult. Hence wider experimental studies of the scattering form-factor would provide a significant benefit to the analysis method that has been outlined here. 
  
A further factor in the success of the present work may be a fortuitous choice of the adsorbate coverage in relation to the strength of the interactions and the surface temperature.  The low coverage and the strong pairwise forces together ensure that the quasi-elastic contribution to the structure factor deviates from unity only at small values of $\Delta \mathbf{K}\leq 0.6\,\mbox{\AA}^{-1}$ as in figure \ref{fig:fig2}(c).  Thus the effect of adsorbate interactions are evident in a region of  $\Delta \mathbf{K}$-space that is clearly different from the effects of the shorter range of forces from the substrate potential.  The latter will be most evident at larger $\Delta \mathbf{K}$.  Similarly, the temperature in the present work is low enough to allow strong correlations to emerge in the adsorbate dynamics, which in turn suggests large differences between coherent and incoherent scattering. At higher temperatures the differences between coherent and incoherent will be reduced as the thermal forces dominate the dynamics. It follows that, in the high-temperature limit, it would be difficult to distinguish between thermal forces and pairwise forces from the $\Delta \mathbf{K}$ dependence of the dephasing rates.  In that regime a two-bath model for interactions would then be appropriate\cite{MarinezCasadoJPCM2007,Martinez-CasadoPRL2007}. 
  
Although the derivation of equation \ref{eq:incohalpha} assumed weak forces, which suggests the results should be approximate for the strongly correlated dynamics of sodium diffusing on copper, it is nevertheless remarkably successful. Our results use stronger interactions, applied in a dynamical range with more highly correlated motion than earlier work\cite{Leitner2011} and they support the  suggestion that the method (\autoref{eq:incohalpha}) has a wider application than is implied by the approximation used in its derivation. 

\section*{Conflicts of interest}
There are no conflicts to declare

\section*{Acknowledgements}
The authors acknowledge use of and support by the \href{https://atomscattering.phy.cam.ac.uk}{Cambridge Atom Scattering Facility} and EPSRC award EP/T00634X/1.  One of us (A.T.) acknowledges financial support provided by the FWF (Austrian Science Fund) within the project J3479-N20. S.M.A. would like to acknowledge the Ministerio de Ciencia, Universidades e Innovacion (Spain) for the grant with Ref. FIS2017-83473-C2-1-P. We thank the Cavendish Workshop, particularly Rik Balsod, for making the titanium dispenser that was critical for taking the experimental data.

\providecommand*{\mcitethebibliography}{\thebibliography}
\csname @ifundefined\endcsname{endmcitethebibliography}
{\let\endmcitethebibliography\endthebibliography}{}

\end{document}